\documentclass[aps,prl,showpacs,twocolumn,superscriptaddress,amsmath,amssymb,floatfix]{revtex4-1}
\usepackage{tabularx}
\usepackage{gensymb,wasysym}
\usepackage{bm}
\usepackage{epsfig,subfigure}
\usepackage{multirow}
\usepackage{graphicx}
\usepackage{color}
\usepackage{amsfonts}
\usepackage{exscale}
\usepackage{amsbsy}
\usepackage{soul}
\graphicspath{{Fig/}}
\usepackage[colorlinks,urlcolor=blue,linkcolor=blue,anchorcolor=blue,citecolor=blue]{hyperref}

\newcommand{\cs}[1]{\varepsilon_{\mu\nu\rho} #1_\mu \partial_\nu #1_\rho}
\newcommand{\scs}[1]{ #1_\mu \partial_\nu #1_\rho}
\newcommand{\vp}{\varepsilon_{\mu\nu\rho}}
\newcommand{\mcs}[2]{\varepsilon_{\mu\nu\rho} #1_\mu \partial_\nu #2_\rho}
\newcommand{\smcs}[2]{#1_\mu \partial_\nu #2_\rho}
\newcommand{\mw}[1]{(\partial_\mu #1_\nu - \partial_\nu #1_\mu)^2}

\begin{document}
\begin{abstract}

We investigate the nature of quantum spin liquid (QSL) phase of the spin-$1/2$ nearest-neighbor XXZ antiferromagnet on the kagome lattice. 
Recent numerical calculations suggest that such a kagome spin liquid (KSL) is insensitive to the XXZ anisotropy and is adjacent to a chiral spin liquid phase. 
Reformulating the problem in terms of a $U(1)$ lattice gauge theory with dynamical bosonic spinons, we propose that the KSL can be understood as a deconfined critical phase extended by the $U(1)$ gauge fluctuation from a deconfined critical point between a symmetry-protected  topological phase and a superfluid phase. 
Crucially, the stability of this QSL is ensured by the gauge fluctuation and hence our description necessarily falls beyond the conventional mean-field constructions of QSLs. 
Our work also makes an interesting connection between the critical spin liquid, deconfined criticality, symmetry protected topological phase and topological order.
\end{abstract}

\title{Kagome spin liquid: a deconfined critical phase driven by $U(1)$ gauge fluctuations}
\author{Yin-Chen He}
\altaffiliation{Current address: Department of Physics, Harvard University, Cambridge, Massachusetts 02138, USA}
\affiliation{Max-Planck-Institut f\"{u}r Physik komplexer Systeme, N\"{o}thnitzer Str. 38, 01187 Dresden, Germany}
\author{Yohei Fuji}
\affiliation{Max-Planck-Institut f\"{u}r Physik komplexer Systeme, N\"{o}thnitzer Str. 38, 01187 Dresden, Germany}
\author{Subhro Bhattacharjee} 
\affiliation{International Centre for Theoretical Sciences, Tata Institute of Fundamental Research, Bangalore 560089, India}
\maketitle

A quantum spin liquid (QSL) \cite{Anderson1973,Wen_book} offers an example \cite{Moessner2001,Balents2002, Kitaev2006, Schroeter2007, Yan2011, He2014, ssgong14, Bauer2014} of long-range quantum-entangled states of condensed matter that necessarily fall beyond Landau's symmetry breaking paradigm of classifying condensed matter phases \cite{Wen_book}. Such phases can support fractional low energy excitations such as spin-1/2 spinons and emergent photons. The nearest-neighbour spin-$1/2$ Heisenberg antiferromagnet on the kagome lattice is a leading candidate  that possibly realise a QSL ground state. Given  its apparent simplicity, this model has been the subject of intense research over the last couple of decades  \cite{Sachdev1992_kagome, Yang1993, Hastings2000, Ran2007, Yan2011, Depenbrock2012, Jiang2012, Iqbal2013}, but, has eluded comprehensive understanding so far. 

A remarkable  insight on this issue was recently obtained from the numerical density matrix renormalization group (DMRG)  calculations of Yan  et. al. \cite{Yan2011}. 
They provided a strong evidence for a QSL ground state in this system. Though, later DMRG works \cite{Depenbrock2012,Jiang2012} suggested it is a $Z_2$ QSL, the characteristic properties of the $Z_2$ QSL, i.e. four-fold topological degeneracy and fractional statistics of spinons, supposedly identifiable  in a straightforward manner using DMRG \cite{Cincio2013,He2014a}, has not been found yet.  Therefore, despite encouraging indications, the precise nature of the enigmatic ``kagome spin liquid" (KSL) still remains an important open problem.

A key reason for the difficulty in understanding the nature of the KSL has been a lack of controlled theoretical methods to directly analyze the microscopic spin model. However, a recent numerical work \cite{He2015} has provided useful clues to a possible theoretical handle.  This involves considering a general class of spin-1/2  kagome antiferromagnets with XXZ anisotropy that includes the Heisenberg model as a special case. It yields the following interesting results:  (1) the KSL in the Heisenberg model is adiabatically connected to the easy-axis as well as the easy-plane limit \cite{He2015, lauchli2015}, and (2) by adding suitable second and third neighbour interactions, a chiral spin liquid (CSL) phase \cite{Kalmeyer1987, Wen1989} (with  spontaneously broken time-reversal symmetry  (TRS)) is obtained, possibly through a continuous phase transition.  Both the CSL and the transition again appear to be independent of the XXZ anisotropy.  These insights have opened up a possible way to understand the KSL from the easy-axis limit, in which one can formulate a faithful lattice gauge theory in terms of a compact $U(1)$ gauge field coupled to dynamical bosonic spinons \cite{Nikolic2005a, He2015c}.  This approach has been already successful in understanding  the CSL \cite{He2015c}, which in this framework turns out to be a ``gauged" \cite{Barkeshli2013, Levin2012} $U(1)$ symmetry protected topological (SPT \cite{Chen2013, Pollmann2010, Vishwanath2013}) phase, the bosonic integer quantum Hall state \cite{Lu2012, Senthil2013}).

In this paper, we use the lattice gauge theory to investigate the nature of the KSL in the easy-axis limit. Our central conclusion is:  the KSL is a deconfined critical phase driven by $U(1)$ gauge fluctuations.  We argue that this deconfined critical phase is best understood through the concept of ``gauging" of a global $U(1)$ symmetry of a deconfined critical point between a bosonic $U(1)$ SPT phase (with  spontaneously broken TRS) and a superfluid phase. Since gauge fluctuations, that promote the critical point to an extended critical phase, are essential for the stability of the phase, our description of the KSL is necessarily beyond the mean-field constructions of QSLs.  We sketch out the above proposal in terms of a QED3 theory (in a similar form as Ref. \cite{Grover2013,Lu2014} ) that describes  the $U(1)$ SPT phase and its deconfined phase transition \cite{Senthil2004a,Motrunich2004,Senthil2006} to a superfluid.
Our theory might also have interesting relation with the recent proposed particle-vortex duality of Dirac fermions \cite{Wang2015a,Wang2015b, Metliski2015,Wang2015c,Geraedts2015,Mross2015b, Xu2015}.

\begin{figure}
\includegraphics[width=0.44\textwidth]{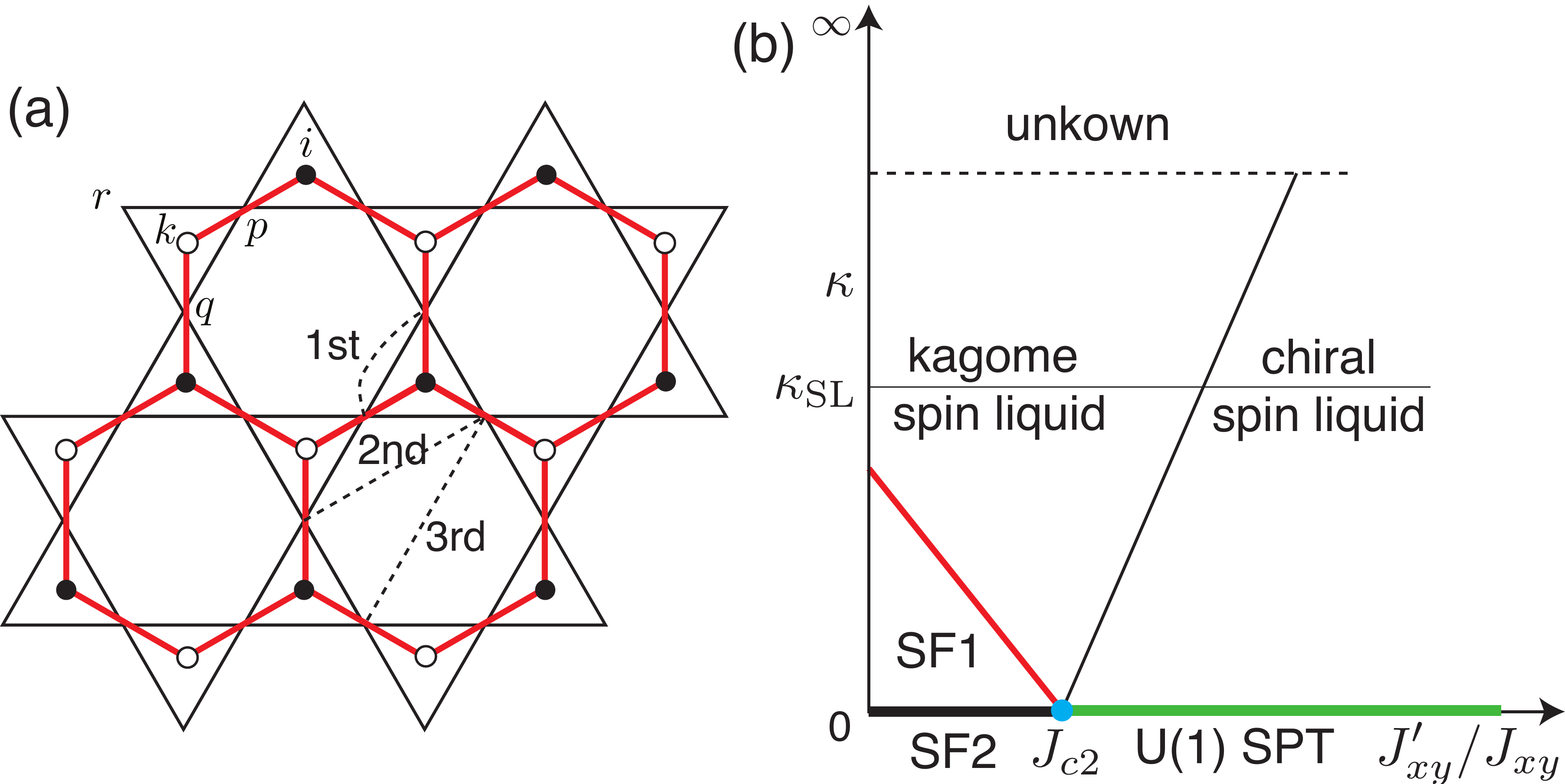}
\caption{ \label{fig:kagome}  (color online) (a) Kagome lattice and the medial honeycomb lattice. The third-neighbor interaction is within the hexagon plaquette. The lattice gauge theory is defined on the medial honeycomb lattice. (b) The proposed phase diagram for the lattice gauge model Eq. (\ref{eq:LGT_Ham}). The $\kappa=\kappa_{\rm SL}$ line corresponds to the physics of easy-axis kagome in the Eq. (\ref{eq:XXZ}), $J_z\gg J_{xy}, J_{xy}'$. 
The $\kappa=0$ line corresponds to the Eq. (\ref{eq:boson_Ham}). SF1 (SF2) refer to  superfluid phases with one (two) Goldstone mode(s). When $\kappa$ is large, the system might enter some new phase, which however is beyond the scope of present work.}
\end{figure}

We begin with an extended XXZ kagome model \cite{He2015}, 
\begin{align}\label{eq:XXZ}
H_{XXZ} &=J_z \sum_{\langle p q\rangle} S_p^z S_q^z+\frac{J_{xy}}{2}\sum_{\langle p q\rangle}  (S^+_p S^-_q +h.c.) \nonumber  
\\ &+ \frac{J'_{xy}}{2}\Big(\sum_{\langle\langle p q\rangle\rangle}+\sum_{\langle\langle\langle p q\rangle\rangle\rangle}\Big) (S^+_p S^-_q +h.c.),
\end{align}
where $\langle pq \rangle$ denotes nearest neighbour XXZ interactions and $\langle\langle pq \rangle\rangle$ and $\langle\langle\langle pq \rangle\rangle\rangle$ denote second and third neighbour XY interactions (see Fig. \ref{fig:kagome}(a)). The DMRG calculations on this model has found two QSL phases, the KSL and the TRS broken CSL. A surprising feature of the numerically obtained phase diagram is its independence of the XXZ anisotropy, namely the two QSL phases extend from the easy-axis limit ($J_z\gg J_{xy}>0$) through the Heisenberg point ($J_z=J_{xy}$), to the easy-plane limit ($J_z=0, J_{xy}>0$) with the transition between them  likely being a continuous one throughout.

In the easy-axis limit ($J_z\gg J_{xy}, J_{xy}'$), the QSL phases can be studied using a lattice gauge theory\cite{He2015c,Nikolic2005a}. The mapping follows by noting that, in this limit,  strong nearest neighbour Ising interactions constrain the system to live in the classically degenerate manifold that fulfills $\sum_{p \in \bigtriangleup,\bigtriangledown} S_p^z=\pm 1/2$ for each triangle of the kagome lattice.  Then one can define two flavors of hard-core bosonic spinons that live at the center of the up/down triangles (which forms the two sub-lattices of the medial honeycomb lattice) as $\sum_{ p \in \bigtriangleup_i} S_{ p}^z  =a^\dag_i a_i- 1/2$,   $\sum_{ p \in \bigtriangledown_k} S_p^z =b^\dag_k b_k- 1/2$.  Here $a^\dag$, $b^\dag$ respectively refer to spinon creation operators on the A, B sublattices of the honeycomb lattice.  %The spinons satisfy the hard-core constraint. 
The spin-flip operators is written in terms of the spinons as $S^+_p = \exp (i \mathcal A_{ik})  a_i^\dag  b_k^\dagger$. The $\mathcal A_{ij}\in [0,2\pi)$ are compact $U(1)$ gauge fields living on the links of the honeycomb lattice and the spinons are minimally coupled to this gauge field. 
The electric field, which is conjugate to the above gauge field, is given by $E_{ik}=- E_{ki}=(S^z_{ p}+1/2)$. 

Using the above mapping, the effective Hamiltonian for Eq. \eqref{eq:XXZ} in the easy-axis limit becomes, 
\begin{align} \label{eq:LGT_Ham}
H^{\rm LGT}&=J_{xy}  \Big[\sum_{\langle \langle ij \rangle\rangle}e^{i \mathcal  A_{ij}}  a^\dag_i a_j + \sum_{\langle \langle kl \rangle\rangle} e^{i\mathcal A_{lk}} b^\dag_k b_l + h.c. \Big] \nonumber
 \\ & + J'_{xy} \sum_{\langle i k\rangle, \langle j l \rangle \in \hexagon}  \left[(e^{i\mathcal A_{ik} } a^\dag_i b_k^\dag)( e^{ i\mathcal A_{lj}} b_l a_j) +h.c.\right] \nonumber \\
& + \kappa\sum E_{ik}^2+ \frac{1}{\kappa} \sum \cos(\sum_{\hexagon} \mathcal A_{ik}).
\end{align}
The first two lines describe the dynamics of the bosonic spinons coupled to the compact $U(1)$ gauge field while the last line describes the dynamics of the gauge field (Maxwell term).  The coupling constant $\kappa$  represents the strength of gauge fluctuation. The physics of the easy-axis kagome antiferromagnet corresponds to some  $\kappa=\kappa_{\rm SL}\neq 0$, whose exact magnitude is not important for the purpose of this paper. However the sign of $\kappa$ is important and in our regime of interest ($J_{xy}, J_{xy}'>0$), $\kappa>0$ and this leads to trapping of $\pi$ flux, through each hexagonal plaquette.  So, we outline a general phase diagram for this lattice gauge model in Fig. \ref{fig:kagome}(b) as $\kappa (>0)$ and $J_{xy}'/J_{xy}$ varies. This phase diagram is the central result of this work and the rest of the paper is devoted in explaining this phase diagram, emphasizing the physics of the KSL. 

We start by drawing attention to two special lines of the phase diagram, $\kappa=\kappa_{\rm SL}$ and $0$, which correspond to two interesting microscopic models. $\kappa=\kappa_{\rm SL}$, of course, corresponds to the easy-axis kagome model,  which hosts the KSL for $J'_{xy}/J_{xy}\in [0, J_c)$, the CSL for $J'_{xy}/J_{xy} \in (J_c, 2]$, and a possible continuous transition point at $J_c\approx 0.6$ \cite{He2015}. On the other hand, $\kappa=0$ corresponds to the case where the gauge fluctuations are frozen to a static background flux $A^0$ ($ \sum_{\hexagon}  A^0=\pi$), yielding
\begin{align} \label{eq:boson_Ham}
H_{\textrm{b}}&=J_{xy}  \Big[\sum_{\langle \langle ij \rangle\rangle}e^{i   A^0_{ij}}  a^\dag_i a_j + \sum_{\langle \langle kl \rangle\rangle} e^{i A^0_{lk}} b^\dag_k b_l + h.c. \Big] \nonumber
 \\ & + J'_{xy} \sum_{\langle i k\rangle, \langle j l \rangle \in \hexagon}  \left[(e^{i A^0_{ik} } a^\dag_i b_k^\dag)( e^{ i A^0_{lj}} b_l a_j) +h.c.\right].
 \end{align}

This Hamiltonian describes two species of boson with two global $U(1)$ symmetries, $U(1)_c$ charge and $U(1)_s$ pseudospin; $a^\dag$ carries pseudospin up and $b^\dag$ carries pseudospin down and both carries same charge.
For  $J'_{xy}=0$ in Eq. \ref{eq:boson_Ham}, this is nothing but hard-core bosons hopping within the second-neighbour links of the honeycomb lattice in a background of $\pi$ flux, resulting in a superfluid.
On the other hand, for $J_{xy}'/J_{xy}\in (J_{c2}, 1.0]$ ($J_{c2}\approx 0.2$), it  hosts a $U(1)$ SPT phase with  spontaneous breaking of TRS \cite{He2015c}. Moreover it is likely, as we will show, that the superfluid and the $U(1)$ SPT phase are separated by a direct continuous phase transition.

Having described the physics expected on the $\kappa=\kappa_{\rm SL}$ and $\kappa=0$ lines, we now turn to the question of the general structure of the phase diagram and more specifically, how the physics on these two special lines are connected. Previous work \cite{He2015c} shed light on this connection for the CSL phase, which is obtained by gauging a global $U(1)$ symmetry of the $U(1)$ SPT phase, which, in the specific context of Eq. (\ref{eq:boson_Ham}), is the pseudospin $U(1)_s$. In other words, turning on a finite gauge fluctuations $\kappa$ in the $U(1)$ SPT phase (on $\kappa=0$ line) results in the CSL phase that survives for $\kappa=\kappa_{\rm SL}$ (with the phase boundary modified by the gauge fluctuation). 

On the other hand, for $\kappa=0$ and $J'_{xy}=0$, we have a superfluid phase that spontaneously breaks $U(1)_c\times U(1)_s$ and hence has two Goldstone modes. 
Infinitesimal gauge fluctuations ($\kappa\neq 0$), then, will Higgs out the Goldstone mode corresponding to the $U(1)_s$ breaking, yielding a superfluid with one Goldstone mode for the breaking of $U(1)_c$. This is nothing but an XY magnetic ordered state for the spin model. 
As the gauge fluctuation further increases to critical value, ($\kappa_c<\kappa_{\rm SL}$), the superfluid phase will be disordered giving way to the KSL through a phase transition. This KSL, as we shall now show, can then be visualized as an extension of the critical point (at finite $J_{xy}'$) between the superfluid and the $U(1)$ SPT phase at $\kappa=0$, as shown in Fig. \ref{fig:kagome} (b).

\emph{Effective theory at $\kappa=0$.}---At $\kappa=0$ (See Eq. \ref{eq:boson_Ham}), we have two species of boson with the $U(1)_c\times U(1)_s$ symmetry as noted above. By tuning the ratio of $J'_{xy}/J_{xy}$ from $0$ to a finite value in Eq. (\ref{eq:boson_Ham}), the system will transit from the superfluid to the $U(1)$ SPT phase (exhibiting TRS breaking). 
We use a QED3  theory to describe it, it has  two-component Dirac spinors $\psi^f_\pm$, $\psi^g_\pm$ coupled to internal gauge fields $a^f$, $a^g$ respectively.  Thus the resultant theory that we propose, which is also the starting point of our analysis, is
\begin{align} \label{eq:QED3}
 \mathcal L=&\sum_{\sigma=\pm}\bar \psi^f_\sigma [i\gamma^\mu(\partial_\mu-ia^f_\mu-i\sigma A^c_\mu) ]\psi^f_\sigma -\frac{1}{2\pi}\mcs{A^s}{a^f} \nonumber \\
+&\sum_{\sigma=\pm}\bar \psi^g_\sigma [i\gamma^\mu(\partial_\mu-ia^g_\mu-i\sigma A^c_\mu) ]\psi^g_\sigma +\frac{1}{2\pi}\mcs{A^s}{a^g} \nonumber \\
+&\sum_{\sigma=\pm}\phi(\bar \psi^f_\sigma \psi^f_\sigma+\bar \psi^g_\sigma \psi^g_\sigma) +2\lambda \phi^2 -u \phi^4+(\partial_\mu \phi)^2 \nonumber \\
+& \mw{a^{f}}/4e^2+\mw{a^{g}}/4e^2,
\end{align}
where  $A^c$ and $A^s$ are the probe gauge fields that are coupled to the global $U(1)_c$ charge $Q_c$ and $U(1)_s$ pseudospin $Q_s$ respectively. Also,  a bosonic field $\phi$ is coupled to the Dirac fermions.
In the microscopic model, the field $\phi$ is a TRS breaking order parameter corresponding to the correlated hopping $i(2n_k^b-1)a^\dag_i a_j$ for the bosons in Ref. \cite{He2015b, Fuji2016}).
Crucially enough, the above fermionic field theory can be systematically derived starting from a bosonic system,  which may be related with Eq. (\ref{eq:boson_Ham}) via a coupled-wire construction \cite{Fuji2016, Mross2015b} (supplementary section).
A similar theory describing the transition between the $U(1)$ SPT and the Mott insulator was proposed  based on a parton construction  \cite{Grover2013, Lu2014}.

The above field theory can be considered as two-copies of QED3, and under the bosonic TRS : %(complex conjugate $\mathcal K$)
 \begin{equation}
 \psi^f_\sigma \rightarrow i\tau_y (\psi^g_\sigma), ~~~ a^f_0 \rightarrow a^g_0, ~~~ a^f_i \rightarrow -a^g_i, ~~~ \phi \rightarrow -\phi,
 \end{equation}
these two copies are exchanged to each other, making the Lagrangian TRS invariant ($\tau_y$ acts on the internal spinor space.). 
Also, the above field theory is not anomalous and can be realized in pure two (spatial) dimension, unlike the theory of surface state of 3D bosonic SPT. 

Similar to the Son's Dirac-composite-fermion theory, the Dirac fermions in our theory are the vortices of the $U(1)_s$ pseudospin. One quick way to understand this fact is to realize that the pseudospin current is
$j^s_\mu=\frac{1}{2\pi} \vp \partial_\nu(a^f_\rho-a^g_\rho)$.

In general, $\lambda$ controls the low energy physics, namely when $\lambda>0$, the system is in a $U(1)$ SPT phase with spontaneous TRS breaking; when $\lambda<0$, the system is in a superfluid phase.
$\lambda=0$ is the critical point, at which the gapless Dirac fermions are carrying fractional charge, $Q_c=\pm 1$ and $Q_s=0$. 
But as we will show below, this pseudospin-charge separation phenomena only happens at the critical point (that's why it is a deconfined critical point), not in the superfluid or the $U(1)$ SPT phase.

\paragraph{(i)  For $\lambda>0$,} TRS will be spontaneously broken, and $\phi$ takes a finite value $ \phi_0\approx\pm \sqrt{\lambda/u}$. 
As a result, the Dirac fermions will acquire a finite mass $ \phi_0$, hence can be integrated out and yield (Maxwell terms are omitted) 
 \begin{align}
&\mathcal L=-\frac{\textrm{sgn} (\phi_0)}{4\pi} \vp[ \scs{A^c}+\scs{a^f}]-\frac{\vp}{2\pi}\smcs{A^s}{a^f} \nonumber \\
		  & -\frac{\textrm{sgn} (\phi_0)}{4\pi} \vp[\scs{A^c}+\scs{a^g}]+\frac{\vp}{2\pi}\smcs{A^s}{a^g}  		 % &+\frac{1}{4e^2} \mw{a^f}+\frac{1}{4e^2} \mw{a^g}.
\end{align}
Due to the emergence of Chern-Simons terms for $a^f$ and $a^g$, the photons of the gauge fields $a^f$, $a^g$ acquire a mass $\sim e^2$ and hence are gapped.
Therefore, even though the vortices (Dirac fermions) are gapped here, the system is  in an insulating phase. Another important point is, due to the Chern-Simons terms, the fermionic vortex will be dressed up with one Chern-Simons flux quanta and thus changes its statistics from  fermionic to bosonic.   Also there is no pseudospin-charge separation anymore, since the flux quanta of $a^f$ (or $a^g$) will  bind pseudospin $Q_s=1$ to the Dirac fermions.
Integrating out $a^f$ and $a^g$ gives
\begin{align}
\mathcal L=&-\frac{2}{4\pi}\textrm{sgn} (\phi_0) \vp[ \scs{A^c}-\scs{A^s}].
\end{align}
Therefore, we will end up with a $U(1)$ SPT phase with charge Hall conductance $\sigma^{xy}_c=2\,\textrm{sgn} (\phi_0)$.
\paragraph{(ii) For $\lambda<0$,} $\phi$ is gapped, and can be safely integrated out. The resultant theory is nothing but two copies of QED3 with $N_f=2$ flavors of Dirac fermions, which is unstable to, for example, spontaneous chiral symmetry breaking with the generation of a mass term,
$ \mathcal L_{\rm CSB}= \sum_{\sigma=\pm} \sigma( m^f \bar \psi^f_\sigma \psi^f_\sigma+ m^g \bar \psi^g_\sigma \psi^g_\sigma)$.
Here $m^f$, $m^g$ represent some finite masses whose specific values depend on microscopic details.
Similar as before, we integrate out those gapped Dirac fermions (say $m^f, m^g>0$) , 
\begin{align}
\mathcal L=-\frac{\vp}{2\pi}\smcs{A^c}{(a^f+a^g)}+\frac{\vp}{2\pi}\smcs{A^s}{(a^f-a^g)}
\end{align}
This is the superfluid phase (not the anyon superfluid in Ref. \cite{Grover2013, Lu2014}), since the vortices are gapped and the gapless photons of the gauge fields $a^{f(g)}$ correspond to the Goldstone modes of the superfluid \cite{boson-vortex}.  Here we have two Goldstone modes from the $U(1)_c$ charge and $U(1)_s$ pseudospin symmetry breaking respectively. There is no longer pseudospin-charge separation in this superfluid phase as the charge is bound with pseudospin (can be seen by integrating out $a^{f(g)}$).
\paragraph{(iii) $\lambda=0$} is the critical point, at which the gapless Dirac fermions are carrying fractional charge, $Q_c=\pm 1$ and $Q_s=0$.   
At $\lambda=0$, we have two copies of $N_f=2$ massless QED3, but in contrast to the $\lambda<0$ case, those Dirac fermions are coupled to a gapless bosonic field $\phi$. We expect that due to the coupling with the gapless bosonic field $\phi$, the instability (e.g. the chiral symmetry breaking) might be suppressed, resulting in a stable deconfined critical point.

\emph{Deconfined critical phase extended by gauge fluctuation.}---With the effective theory \eqref{eq:QED3} at $\kappa=0$, we are now ready to attack the finite $\kappa$ case. At the level of effective field theory, finite $\kappa$ corresponds to the gauged version of the Dirac theory in Eq. (\ref{eq:QED3}), namely the global $U(1)_s$ pseudospin symmetry is promoted into a local gauge structure.  This can be easily implemented in Eq. (\ref{eq:QED3}) by replacing the probe gauge field $A^s$ with an internal compact $U(1)$ gauge field $\mathcal A$ and adding a Maxwell term for it.
Meanwhile, we also replace the probe gauge field $A^c$ with $A^{\rm ext}/2$, and $A^{\rm ext}$ is coupled to the quantum number $S^z$ of the spin-1/2 kagome antiferromagnet \cite{He2015c}. Under this ``gauging" process, the $U(1)$ SPT phase on the  $\kappa=0$ line becomes a CSL phase as described in Ref.  \cite{He2015c}. The superfluid with two Goldstone modes at $\kappa=0$, as stated before, will be gauged to a superfluid with one Goldstone mode corresponding to $U(1)_c$.  The gauged version of the critical theory is therefore given by  
$\mathcal L =\mathcal L_D+\mathcal L_{a, \mathcal A}$,
\begin{align}
 \mathcal L_{D}&=\sum_{\sigma=\pm}\bar \psi^f_\sigma [i\gamma^\mu(\partial_\mu-ia^f_\mu-i\sigma \frac{A^{\rm ext}_\mu}{2} ]\psi^f_\sigma \nonumber \\
 &+\sum_{\sigma=\pm}\bar \psi^g_\sigma [i\gamma^\mu(\partial_\mu-ia^g_\mu-i\sigma \frac{A^{\rm ext}_\mu}{2} ]\psi^g_\sigma
 \label{eq:gauged_theory2}
  \\
\mathcal L_{a, \mathcal A}&= \frac{1}{2\pi} \mcs{\mathcal A}{(a^g-a^f)} +\frac{1}{4e_0^2} \mw{\mathcal A} \nonumber
 \\ &+\frac{1}{4e^2} \mw{a^f}+\frac{1}{4e^2} \mw{a^g}.  \label{eq:gauged_theory3}
\end{align}

This is nothing but a gauged version of the two copies of $N_f=2$ QED3, which now might describe an extended critical phase. 
To see why this is the case, it is more straightforward to consider the regime $e_0>e$ (but $e_0\ll \infty$),
in which we can simply integrate out $\mathcal{A}$, yielding a mass for the internal gauge fields of the form $\sim e_0^2 (a^f-a^g)^2$ and hence locking $a^f=a^g$. For $e_0<e$, the physics is similar (see the supplementary materials). Due to this locking, the two copies of $N_f=2$ QED3 gets converted into a single $N_f=4$ QED3,
\begin{align}\label{eq:final}
&\mathcal L=\sum_{\substack{\sigma=\pm\\ v=f,g}}\bar \psi^v_\sigma [i\gamma^\mu(\partial_\mu-ia_\mu-i\sigma \frac{A^{\rm ext}_\mu}{2}) ]\psi^v_\sigma+\cdots %\nonumber \\
\end{align}

Arguments from large $N_f$ gauge theories coupled to Dirac fermions \cite{Hermele2004b, Grover2014} suggest that for $N_f=4$, such a theory might lead to a stable algebraic spin liquid phase. 
One should contrast the above possibility with the fate of the ``un-gauged" theory in Eq. (\ref{eq:QED3}), which has (two copies) $N_f=2$ and hence the critical state is  fine tuned and is restricted to a single point. Thus the gauge fluctuation (of $\mathcal A$) is  necessary to open up a critical phase.
Compared with the previous Dirac spin liquid in Refs. \cite{Hastings2000,Ran2007,Iqbal2013}, our theory has other interactions, for instance some term that explicitly breaks the $SU(4)$ symmetry. 
As a future direction, it would be very interesting to understand the infrared properties of such a field theory and contract it with the properties of the Dirac QSL obtained from the parton constructions.

Alert readers may realize that there is a caveat that, to reach Eq. (\ref{eq:final}), we  integrate out the compact $U(1)$ gauge field $\mathcal A$. 
The compactness of $\mathcal A$ is actually important, as it might cause confinement by proliferating monopoles (instantons)  in $2+1$ dimension \cite{Polyakov}.
However, note  the monopoles and the anti-monopoles are indeed associated with the vortex rings \cite{Einhorn1978}, which are our Dirac fermions $\psi^f_\pm,\psi^g_\pm$. These fields being gapless generate long range interactions between monopoles and anti-monopoles, binding them to form magnetic dipoles and thus suppressing the confinement caused by the compactness of $\mathcal A$. 
 It is interesting to know the relation between the physical operators (e.g. spin) and fermionic field.
Such relation can be obtained using a coupled wire approach, which however is very technical and beyond the scope of present work.
But based on symmetry, we can fix the form of this relation as 
$\vec S \sim (\psi^{f})^\dag \vec \sigma \psi^f + (\psi^{g})^\dag \vec \sigma \psi^g$,
and the scalar chirality is
$\chi=\vec S_p \cdot (\vec S_q \times \vec S_r) \sim c_1 \phi+ c_2 (\bar \psi^f \psi^f +\bar \psi^g \psi^g)$,
where $c_1$ and $c_2$ are some number.

\emph{Conclusion and outlook} --- We investigate  the nature of the kagome spin liquid (KSL) realized in the easy-axis XXZ spin-1/2 kagome antiferromagnets  using a lattice gauge description consisting of $U(1)$ gauge field coupled to dynamic bosonic spinons. 
Starting with an effective field theory for the deconfined critical point between a superfluid and a $U(1)$ SPT phase (with spontaneous TRS breaking), we show that the KSL can be obtained by promoting a global $U(1)$ symmetry of the bosons to a local gauge structure.  In our approach the gauge fluctuation converts the fine-tuned deconfined critical point into an extended critical phase. In this sense the KSL is obtained by ``gauging" the deconfined critical point. 
This is a concrete implementation of the perceived deep connections among SPTs, associated critical points, and QSLs, through the concept of gauging global symmetries.

It is interesting to compare our theory with the $U(1)$ Dirac spin liquid  (DSL) from parton constructions \cite{Hastings2000,Ran2007,Iqbal2013}.
Due to the inherent importance of the gauge fluctuations, our approach is beyond parton based mean-field construction of QSLs. 
However, presently, it is unclear whether our QSL phase is  different from the previous DSL  or not.  One way to understand the issue is to compare the quantum numbers of the low energy excitations in both cases. 
Even if  they are the same phase, our work still advances the understanding on this celebrated problem by providing a complimentary insight to the problem and more importantly explaining the natural proximity of the CSL phase in this class of models. 
It is also  interesting to ask whether the speculated vortex duality for Dirac fermions \cite{Son2015, Metliski2015,Wang2015b, Wang2015c,Xu2015, Geraedts2015,Mross2015b} could apply for our kagome spin liquid theory.
Such possible duality can be understood as the self-dual of the (ungauged) deconfined critical point, which might further serve as an interesting probe for our theory both  numerically and experimentally.

\emph{Acknowledgements. --}We thank R. Moessner and F. Pollmann for the collaboration on related projects.  YCH  thanks the discussions with L. Motrunich, S. Sachdev, T. Senthil,  A. Vishwanath, C. Wang and M. Zaletel. SB  thanks Visitors Program at MPIPKS for hospitality and  A. Dhar and S. Rao for discussions.

\newpage

\appendix
\widetext

\section{Dirac composite fermion theory from coupled wires}

We here provide a partial derivation of the Dirac composite fermion theory presented in the main text by using the coupled-wire construction. 
Specifically, we start with a  pure bosonic system described by a two-dimensional array of two-component Luttinger liquid,  
\begin{align} \label{eq:network1}
S[\varphi^\pm_j, \tilde{\varphi}^\pm_j, \phi_j] &= S_\textrm{wire}[\varphi^\pm_j, \tilde{\varphi}^\pm_j] +S_\textrm{tun}[\varphi^\pm_j, \tilde{\varphi}^\pm_j, \phi_j], \\
S_\textrm{wire} [\varphi^\pm_j, \tilde{\varphi}^\pm_j] 
&= \int dt dx \sum_{j=0}^{N-1} \Biggl[ \frac{1}{4\pi} \sum_{\sigma,\sigma'=\pm} \sigma^x_{\sigma \sigma'} \Bigl( \partial_x \varphi^\sigma_j \partial_t \varphi^{\sigma'}_j -\partial_x \tilde{\varphi}^\sigma_j \partial_t \tilde{\varphi}^{\sigma'}_j \Bigr) 
-\frac{v}{4\pi} \sum_{\sigma=\pm} \Bigl\{ (\partial_x \varphi^\sigma_j)^2 +(\partial_x \tilde{\varphi}^\sigma_j)^2 \Bigr\} \Biggr], \\
\label{eq:Tunneling}
S_\textrm{tun} [\varphi^\pm_j, \tilde{\varphi}^\pm_j,\phi_j] 
&= -\int dt dx \sum_{j=0}^{N-1} \biggl[ (g_1 -g_2 \phi_j) \sum_{\sigma=\pm} \cos (\varphi^\sigma_j-\tilde{\varphi}^\sigma_{j+1})
+(g_1 +g_2 \phi_j) \sum_{\sigma=\pm} \cos (\tilde{\varphi}^\sigma_j-\varphi^\sigma_{j+1}) \nonumber \\
&\hspace{75pt} -c_1 \phi_j^2 +c_2 (\phi_j-\phi_{j+1})^2 +\cdots \biggr]. 
\end{align}
Applying an explicit duality transformation proposed in Ref.~\cite{Mross2015b} to the coupled-wire model, we can controllably derive a spatially anisotropic version of the QED3 theory which we introduce in the main text.
\begin{align}
S_\textrm{dual}[\psi^{f/g}_\pm, A^{c/s}_\mu, a^{f/g}_\mu, \phi] \sim& \ \int dt dx dy \Biggl[ \sum_{\sigma=\pm} \bar{\psi}^f_\sigma  iw_\mu \gamma^\mu (\partial_\mu-ia^f_\mu -i\sigma A^c_\mu) \psi^f_\sigma +\sum_{\sigma=\pm} \bar{\psi}^g_\sigma iw_\mu \gamma^\mu (\partial_\mu -ia^g_\mu -i\sigma A^c_\mu) \psi^g_\sigma \nonumber \\
&+4\pi g_2 d_x \phi \sum_{\sigma=\pm} (\bar{\psi}^f_\sigma \psi^f_\sigma +\bar{\psi}^g_\sigma \psi^g_\sigma) +\frac{c_1}{d_y} \phi^2 -c_2 d_y (\partial_y \phi)^2 +\frac{1}{2\pi} \epsilon_{\mu \nu \rho} A^s_\mu \partial_\nu (a^f_\rho -a^g_\rho) \nonumber \\
&+\frac{1}{8\pi} \lambda_\mu (\epsilon_{\mu \nu \rho} \partial_\nu a^f_\rho)^2 +\frac{1}{8\pi} \lambda_\mu (\epsilon_{\mu \nu \rho} \partial_\nu a^g_\rho)^2 +\frac{1}{4\pi} \lambda_\mu (\epsilon_{\mu \nu \rho} \partial_\nu A^s_\rho)^2 +\cdots \Biggr], 
\end{align}
Before going to the details, we hope to give several remarks:
\begin{enumerate}
\item The network model given by Eqs. \eqref{eq:network1}-\eqref{eq:Tunneling} can be regarded as two copies of a network model that has been proposed to describe a gapless surface state of the (3+1)-dimensional bosonic topological insulator \cite{Vishwanath2013}.
 Besides, we  couple it with an order-parameter field $\phi_j(t,x)$ associated with time-reversal symmetry breaking, which allows us to describe the phase transition between a superfluid and $U(1)$ SPT phase. 
\item A very similar network model, just with a modification that the real scalar field $\phi_j(t,x)$ in Eq. \eqref{eq:Tunneling} is replaced by a constant, has been shown to exactly describe the low energy physics of a two-component interacting bosonic lattice model \cite{Fuji2016}.
That lattice model also has interesting connection with the two-component bosonic lattice model of the main text, as pointed out in Ref. \cite{He2015c}.
\item The real scalar fields $\varphi^\pm_j(t,x)$ and $\tilde{\varphi}^\pm_j(t,x)$ are physically related with the two-component bosons of the lattice Hamiltonian in the main text. 
For the model in Ref.~\cite{Fuji2016} (which might not apply here), those fields are defined as $\varphi^+_j=\varphi^a_j + \theta^b_j$, $\varphi^-_j=\varphi^b_j + \theta^a_j$, $\tilde{\varphi}^+_j=\varphi^a_j - \theta^b_j$, and $\tilde{\varphi}^-_j=\varphi^b_j -\theta^a_j$, up to zero-mode contributions (Klein factors) to correctly account for the commutation relations. 
Where, $\varphi_j^{a/b}$ and $\theta_j^{a/b}$ are the dual bosonic fields of the Luttinger liquid for two species of bosons $a$ and $b$ on the $j$-th wire. 
\end{enumerate}

\subsection{Coupled-wire action}

We first explain the coupled-wire action given by Eqs. \eqref{eq:network1}-\eqref{eq:Tunneling}. 
The variables $\varphi^\pm_j(t,x)$, $\tilde{\varphi}^\pm_j(t,x)$, and $\phi_j(t,x)$ are real scalar fields. 
$v$, $g_{1,2}$, and $c_{1,2}$ are some constants dependent on the microscopic details. 
The ellipsis in Eq.~\eqref{eq:Tunneling} includes the kinetic terms of $\phi_j$, namely $c_3 (\partial_t \phi_j)^2+c_4 (\partial_x \phi_j)^2$, and also higher-order perturbations such as $\phi^4_j$, $\phi^2_j \phi^2_{j+1}$, and $\cos (\varphi^\sigma_j +\tilde{\varphi}^\sigma_j -\varphi^\sigma_{j+1} -\tilde{\varphi}^\sigma_{j+1})$, which are neglected in the following analysis. 
We have denoted the Pauli matrices as $\sigma^{x,y,z}$: 
\begin{align}
\sigma^x = \left( \begin{array}{cc} 0 & 1 \\ 1 & 0 \end{array} \right), \hspace{10pt}
\sigma^y = \left( \begin{array}{cc} 0 & -i \\ i & 0 \end{array} \right), \hspace{10pt}
\sigma^z = \left( \begin{array}{cc} 1 & 0 \\ 0 & -1 \end{array} \right). 
\end{align} 
Each wire now consists of two pairs of dual bosonic fields, $\varphi_j^\pm(t,x)$ and $\tilde{\varphi}^\pm_j(t,x)$, which satisfy the equal-time commutation relations, 
\begin{align}
[\partial_x \varphi^\sigma_j(x), \varphi^{\sigma'}_{j'}(x')] = -[\partial_x \tilde{\varphi}^\sigma_j(x), \tilde{\varphi}^{\sigma'}_{j'}(x')] = 2i\pi \sigma^x_{\sigma \sigma'} \delta_{jj'} \delta (x-x'). 
\end{align}
These commutation relations are satisfied by choosing, 
\begin{align} \label{eq:BosonComm}
\begin{split}
[\varphi^+_j(x), \varphi^-_{j'}(x')] &= 2i\pi \delta_{jj'} (\Theta(x-x')-1) -2i\pi (1-\delta_{jj'}) \Theta (j'-j), \\
[\tilde{\varphi}^+_j(x), \tilde{\varphi}^-_{j'}(x')] &= -2i\pi \delta_{jj'} \Theta (x-x') -2i\pi (1-\delta_{jj'}) \Theta (j'-j), \\
[\varphi^+_j(x), \tilde{\varphi}^-_{j'}(x')] &= -2i\pi (1-\delta_{jj'}) \Theta (j'-j), \\
[\tilde{\varphi}^+_j(x), \varphi^-_{j'}(x')] &= 2i\pi \delta_{jj'} +2i\pi (1-\delta_{jj'}) \Theta (j-j'), 
\end{split}
\end{align}
while the other commutators are chosen to vanish. 
Here $\Theta(x)$ is the step function, 
\begin{align}
\Theta(x) = \begin{cases} 1 & (x>0) \\ 0 & (x<0) \end{cases}. 
\end{align} 
The above commutation relations ensure the bosonic statistics of the particle operators $e^{i\varphi^\pm_j}$ and $e^{i\tilde{\varphi}^\pm_j}$. 
Under the global charge and pseudospin $U(1)$ symmetries, the fields transform as 
\begin{align} \label{eq:U1Sym}
\begin{array}{lll}
U(1)_c: & \varphi^\sigma_j \to \varphi^\sigma_j +\alpha_c, & \tilde{\varphi}^\sigma_j \to \tilde{\varphi}^\sigma_j +\alpha_c, \\
U(1)_s: & \varphi^\sigma_j \to \varphi^\sigma_j +\sigma \alpha_s, & \tilde{\varphi}^\sigma_j \to \tilde{\varphi}^\sigma_j +\sigma \alpha_s, 
\end{array}
\end{align}
with arbitrary real numbers $\alpha_{c/s} \in [0,2\pi)$. 
Such operators $e^{i\varphi^\pm_j}$ and $e^{i\tilde{\varphi}^\pm_j}$ can be regarded as the bosonic particle operators of one species dressed by the density fluctuations of another species, which naturally emerge in certain interacting systems of two-component bosons \cite{Fuji2016}. 
Since the cosine terms in $S_\textrm{tun}$ preserve both of these $U(1)$ symmetries, they can be naturally identified as the tunnelings of those bosonic operators from neighboring wires. 
They are further coupled with the real scalar field $\phi_j(t,x)$ corresponding to the order parameter of time-reversal symmetry breaking, which might be generated by integrating out the high-energy degrees of freedom. 
Under the time-reversal symmetry, the fields transform as 
\begin{align}
\varphi^\sigma_j \to -\tilde{\varphi}^\sigma_j, \hspace{10pt}
\tilde{\varphi}^\sigma_j \to -\varphi^\sigma_j, \hspace{10pt}
\phi_j \to -\phi_j, 
\end{align}
which leave the action invariant. 
In the following, we take the number of wires $N$ to be an even integer. 

We then consider possible phases of our coupled-wire model. 
Let us suppose $c_2>0$ in the following analysis. 
At the mean-field level, the field $\phi_j$ acquires a finite expectation value $\langle \phi_j \rangle = \pm \langle \phi \rangle \neq 0$ for $c_1>0$ so that the time-reversal symmetry is spontaneously broken. 
Once the fluctuation of $\phi_j$ is neglected, the tunneling terms are relevant perturbations with the scaling dimension 1. 
If $\langle \phi \rangle <0$, the coupling constant of the first term in Eq.~\eqref{eq:Tunneling} becomes larger than that of the second one and will flow to the strong-coupling limit.  
Under the open boundary condition $\varphi^\sigma_N = \tilde{\varphi}^\sigma_N=0$, the unpaired gapless modes $\tilde{\varphi}^\sigma_0$ and $\varphi^\sigma_{N-1}$ are left at the boundaries while the bulk is gapped. 
These edge states are nothing but those expected for the $U(1)$ SPT phase \cite{Lu2012} and thus the system is in a $U(1)$ SPT phase. 
If $\langle \phi \rangle >0$, the latter tunneling term flows to the strong-coupling limit and leads to another $U(1)$ SPT phase with the opposite chirality. 
Therefore, we conclude that the system is in the $U(1)$ SPT phase with spontaneous time-reversal symmetry breaking for $c_1>0$. 

For $c_1<0$, the expectation value $\langle \phi \rangle$ becomes zero so that the system maintains the time-reversal symmetry. 
In this case, the two tunneling terms compete with each other. 
The system may reside in a critical regime described by the two copies of two-flavor QED3 as shown below or develop some long-range order (superfluid or charge-density wave) driven by higher-order perturbations. 

Coupling with the $U(1)$ probe gauge fields for charge ($A^c_{\mu,j}$) and pseudospin ($A^s_{\mu,j}$), the action becomes 
\begin{align}
S_\textrm{wire} [\varphi^\pm_j, \tilde{\varphi}^\pm_j,A^{c/s}_{\mu,j}] 
=& \ \int dt dx \sum_{j=0}^{N-1} \Biggl[ \frac{1}{4\pi} \sum_{\sigma,\sigma'=\pm} \sigma^x_{\sigma \sigma'} \Bigl\{ \partial_x \varphi^\sigma_j (\partial_t \varphi^{\sigma'}_j -2A^c_{0,j} -2\sigma' A^s_{0,j}) -\partial_x \tilde{\varphi}^\sigma_j (\partial_t \tilde{\varphi}^{\sigma'}_j -2A^c_{0,j} -2\sigma' A^s_{0,j}) \Bigr\} \nonumber \\
& -\frac{v}{4\pi} \sum_{\sigma=\pm} \Bigl\{ (\partial_x \varphi^\sigma_j -A^c_{1,j} -\sigma A^s_{1,j})^2 +(\partial_x \tilde{\varphi}^\sigma_j -A^c_{1,j} -\sigma A^s_{1,j})^2 \Bigr\} \Biggr]. 
\end{align}
In contrast to the case in Ref.~\cite{Mross2016}, there are no (staggered) Chern-Simons terms associated with the gauge anomaly, since each wire now consists of nonchiral bosons. 
For our convenience, we choose the $A^{c/s}_{2,j}=0$ gauge so that the probe gauge fields do not enter the tunneling terms $S_\textrm{tun}$. 

\subsection{Duality transformation}

\begin{figure}
\includegraphics[clip,width=0.6\textwidth]{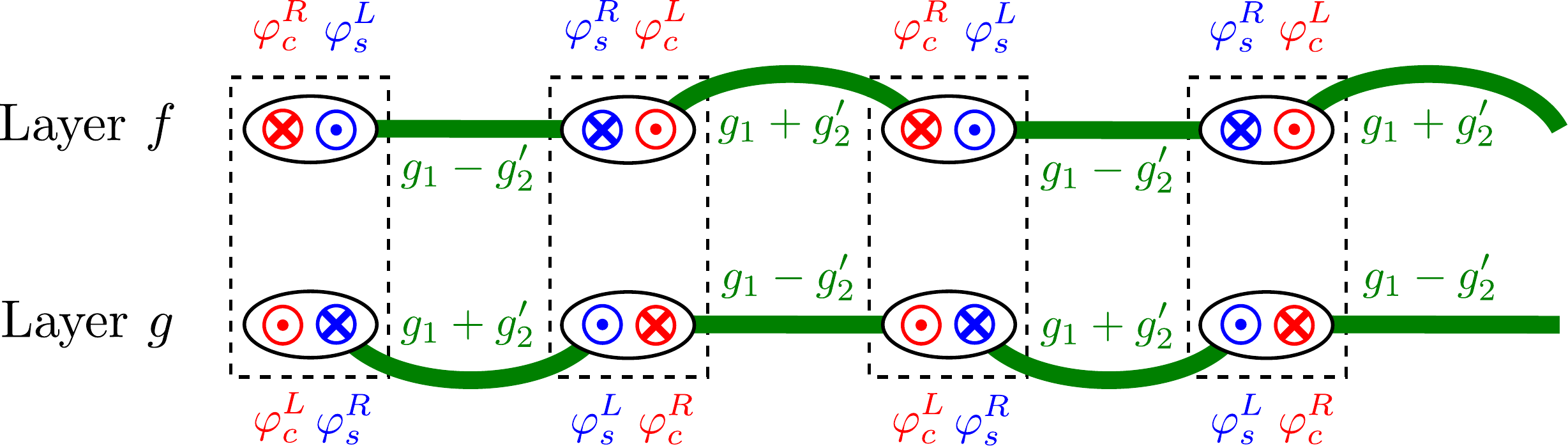}
\caption{Network model view of the action $S_\textrm{wire}+S_\textrm{tun}$. }
\label{fig:TwoLayers}
\end{figure}

Let us introduce the chiral bosonic fields by 
\begin{align}
\begin{split}
\varphi^+_j &= \varphi^R_{c,j} +\varphi^L_{s,j}, \\
\varphi^-_j &= \varphi^R_{c,j} -\varphi^L_{s,j}, \\
\tilde{\varphi}^+_j &= \varphi^L_{c,j} +\varphi^R_{s,j}, \\
\tilde{\varphi}^-_j &= \varphi^L_{c,j} -\varphi^R_{s,j}. 
\end{split}
\end{align}
From Eq.~\eqref{eq:U1Sym}, it is obvious that $\varphi^{R/L}_{c,j}$ carries only charge while $\varphi^{R/L}_{s,j}$ carries only pseudospin. 
Since our coupled-wire model is decoupled into two layers, it is convenient to label the chiral fields in the following way: 
\begin{align}
\begin{split}
(\varphi^f_{c,2n}, \varphi^f_{c,2n+1}) &= (\varphi^R_{c,2n}, \varphi^L_{c,2n+1}), \\
(\varphi^f_{s,2n}, \varphi^f_{s,2n+1}) &= (\varphi^L_{s,2n}, \varphi^R_{s,2n+1}), \\
(\varphi^g_{c,2n}, \varphi^g_{c,2n+1}) &= (\varphi^L_{c,2n}, \varphi^R_{c,2n+1}), \\
(\varphi^g_{s,2n}, \varphi^g_{s,2n+1}) &= (\varphi^R_{s,2n}, \varphi^L_{s,2n+1}), 
\end{split}
\end{align}
where the superscripts $f$ and $g$ refer to the two layers and $n=0,1,\cdots,N/2-1$. 
When $g_2=0$, the $f$ ($g$) layer is described by the same action as that for the top (bottom) surface of the (3+1)-dimensional bosonic topological insulator \cite{Vishwanath2013}. 
However an important difference is that the time-reversal symmetry now transforms one layer to another, 
\begin{align}
\varphi^f_{\rho,j} \to -\varphi^g_{\rho,j}, \hspace{10pt}
\varphi^g_{\rho,j} \to -\varphi^f_{\rho,j} \hspace{10pt}
(\rho=c,s), 
\end{align}
while it transforms back one layer to itself on the surface of the bosonic topological insulator. 
Following Ref.~\cite{Mross2015b}, we introduce the following $N \times N$ matrices, 
\begin{align}
\begin{split}
D_{jk} &= (1-\delta_{jk}) \textrm{sgn}(j-k) (-1)^k, \\
P_{jk} &= (-1)^j \delta_{jk}, \\
\Delta_{jk} &= \delta_{j+1,k} -\delta_{jk}, \\
S_{jk} &= \delta_{j+1,k} +\delta_{jk}, 
\end{split}
\end{align}
as well as the inverse transposition of the lattice derivative $\Delta$, 
\begin{align}
\Delta^{-1,T}_{jk} = \frac{1}{2} \textrm{sgn}(j-k-0^+). 
\end{align}
We also introduce the vector notations, 
\begin{align}
\varphi = (\varphi_0, \cdots, \varphi_{N-1}), \hspace{10pt} (A\varphi) (B\varphi') = \sum_{j,k,l} A_{jk} \varphi_k B_{jl} \varphi'_l, 
\end{align}
where $A$ and $B$ are $N \times N$ matrices. 
In terms of the layer fields $\varphi^{f/g}_{c/s}$, the action is now given by 
\begin{align}
S_\textrm{wire} [\varphi^{f/g}_{c/s}, A^{c/s}_\mu] 
=& \ \int dt dx \biggl[ \frac{P}{2\pi} \Bigl\{ \partial_x \varphi^f_c (\partial_t \varphi^f_c -2A^c_0) -\partial_x \varphi^f_s (\partial_t \varphi^f_s -2A^s_0) 
-\partial_x \varphi^g_c (\partial_t \varphi^g_c -2A^c_0) +\partial_x \varphi^g_s (\partial_t \varphi^g_s -2A^s_0) \Bigr\} \nonumber \\
& -\frac{v}{2\pi} \Bigl\{ (\partial_x \varphi^f_c -A^c_1)^2 +(\partial_x \varphi^f_s -A^s_1)^2 +(\partial_x \varphi^g_c -A^c_1)^2 +(\partial_x \varphi^g_s -A^s_1)^2 \Bigr\} \biggr], \\
S_\textrm{tun} [\varphi^{f/g}_{c/s},\phi_j] =& \ -\int dt dx \sum_{j=0}^{N-1} \sum_{\sigma=\pm} \Bigl[ \big( g_1-(-1)^j g_2 \phi_j \bigr) \cos (\varphi^f_{c,j} +\sigma \varphi^f_{s,j} -\varphi^f_{c,j+1} -\sigma \varphi^f_{s,j+1}) \nonumber \\
&+\big( g_1+(-1)^j g_2 \phi_j \bigr) \cos (\varphi^g_{c,j} +\sigma \varphi^g_{s,j} -\varphi^g_{c,j+1} -\sigma \varphi^g_{s,j+1}) -c_1 \phi_j^2 +c_2 (\phi_j-\phi_{j+1})^2 +\cdots \Bigr]. 
\end{align}
Upon the mean-field treatment, $g_2 \phi_j = g_2 \langle \phi \rangle \equiv g'_2$, the coupling constants in this network model alternate as depicted in Fig.~\ref{fig:TwoLayers}, giving rise to two different $U(1)$ SPT phases depending on the sign of $g'_2$. 
%%%%%%%%%%%%%%%%%%%%%%%%%%%%%%%%%%%%%%%%%%%%%%%%%%%%%%%%%%%%%%%%%
%%%%%%%%%%%%%%%%%%%%%%%%%%%%%%%%%%%%%%%%%%%%%%%%%%%%%%%%%%%%%%%%%

We now apply the duality transformations \cite{Mross2015b} to the chiral pseudospin modes in each layer, 
\begin{align}
\Phi^f_s = D \varphi^f_s, \hspace{10pt} \Phi^g_s = D \varphi^g_s. 
\end{align}
This transformation leaves the tunneling term $S_\textrm{tun}$ local, while the kinetic term involved in $S_\textrm{wire}$ takes a highly nonlocal form. 
In order to make the theory local, we introduce auxiliary fields $a^{f/g}_\mu = (a^{f/g}_0, \cdots, a^{f/g}_{N-1})$ in the path integral and consider the action $S_\textrm{dual} = S_\textrm{wire} +S_\textrm{tun} +S'$ with $S'$ being the complete square form,  
\begin{align}
S'[\Phi^{f/g}_s, A^s_\mu, a^{f/g}_\mu] =& \ -\frac{1}{2\pi} \int dt dx \Biggl[ 2v (\partial_x \Phi^f_s -a^f_1)^2 -v \biggl\{ 2\Delta^{-1,T} P\partial_x \Phi^f_s -\frac{\Delta}{2v} (a^f_0 -DA^s_0 -vDPA^s_1 +vPa^f_1) \biggr\}^2 \nonumber \\
&+2v (\partial_x \Phi^g_s -a^g_1)^2 -v \biggl\{ 2\Delta^{-1,T} P\partial_x \Phi^g_s +\frac{\Delta}{2v} (a^g_0 -DA^s_0 +vDPA^s_1 -vPa^g_1) \biggr\}^2 \Biggr]. 
\end{align}
While the addition of $S'$ turns out to be just the multiplication of a constant factor to the path integral, the resulting action $S_\textrm{dual}$ becomes a local form: 
\begin{align}
&S_\textrm{dual} [\varphi^{f/g}_c, \Phi^{f/g}_s, A^{c/s}_\mu, a^{f/g}_\mu] \nonumber \\
&= S^D_\textrm{wire}[\varphi^{f/g}_c, \Phi^{f/g}_s, A^c_\mu, a^{f/g}_\mu] +S_\textrm{tun}[\varphi^{f/g}_c, \Phi^{f/g}_s, \phi_j] +S_\textrm{mCS}[A^s_\mu, a^{f/g}_\mu]
+S_\textrm {Maxwell}[A^s_\mu, a^{f/g}_\mu] +S_\textrm{stag}[A^s_\mu, a^{f/g}_\mu],
\end{align}
where
\begin{align}
S^D_\textrm{wire}[\varphi^{f/g}_c, \Phi^{f/g}_s, A^c_\mu, a^{f/g}_\mu] 
=& \ \int dt dx \biggl[ \frac{P}{2\pi} \Bigl\{ \partial_x \varphi^f_c (\partial_t \varphi^f_c -2A^c_0) +\partial_x \Phi^f_s (\partial_x \Phi^f_s -2a^f_0) \nonumber \\
&\hspace{10pt} -\partial_x \varphi^g_c (\partial_t \varphi^g_c -2A^c_0) -\partial_x \Phi^g_s (\partial_t \Phi^g_s -2a^g_0) \Bigr\} \nonumber \\
&\hspace{10pt} -\frac{v}{2\pi} \Bigl\{ (\partial_x \varphi^f_c -A^c_1)^2 +(\partial_x \Phi^f_s -a^f_1)^2 +(\partial_x \varphi^g_c -A^c_1)^2 +(\partial_x \Phi^g_s -a^g_1) \Bigr\} \biggr], \\
S_\textrm{tun} [\varphi^{f/g}_c, \Phi^{f/g}_s, \phi_j] =& \ -\int dt dx \sum_{j=0}^{N-1} \sum_{\sigma=\pm} \Bigl[ \big( g_1-(-1)^j g_2 \phi_j \bigr) \cos (\varphi^f_{c,j} +\sigma \Phi^f_{s,j} -\varphi^f_{c,j+1} -\sigma \Phi^f_{s,j+1}) \nonumber \\
&+\big( g_1+(-1)^j g_2 \phi_j \bigr) \cos (\varphi^g_{c,j} +\sigma \Phi^g_{s,j} -\varphi^g_{c,j+1} -\sigma \Phi^g_{s,j+1}) -c_1 \phi_j^2 +c_2 (\phi_j-\phi_{j+1})^2 +\cdots \Bigr], \\
S_\textrm{mCS}[A^s_\mu, a^{f/g}_\mu] =& \ -\frac{1}{4\pi} \int dt dx \Bigl[ \Delta (a^f_0-a^g_0) S A^s_1 +\Delta A^s_0 S (a^f_1-a^g_1) \Bigr], \\
S_\textrm{Maxwell}[A^s_\mu, a^{f/g}_\mu] =& \ \frac{1}{8\pi} \int dt dx \biggl[ \frac{1}{v} (\Delta a^f_0)^2 -v (\Delta a^f_1)^2 +\frac{1}{v} (\Delta a^g_0)^2 -v (\Delta a^f_1)^2 +\frac{2}{v} (\Delta A^s_0)^2 -2v (\Delta A^s_1)^2 \biggr], \\
S_\textrm{stag}[A^s_\mu, a^{f/g}_\mu] =& \ \frac{1}{4\pi} \int dt dx \ P \biggl[ -\Delta a^f_0 S a^f_1 +\Delta a^g_0 S a^g_1 +\frac{1}{v} \Delta (a^f_0+a^g_0) \Delta A^s_0 -v \Delta (a^f_1+a^g_1) \Delta A^s_1 \biggr]. 
\end{align}
Here the actions $S_\textrm{mCS}$ and $S_\textrm{Maxwell}$ for the gauge fields can be regarded as discrete analogues of the mutual Chern-Simons term and Maxwell term, respectively. 
On the other hand, $S_\textrm{stag}$ is rapidly oscillating in the wire index $j$ and thus will be negligible in the continuum limit. 
We hereafter drop this oscillating term. 

Let us introduce the new chiral bosonic fields, 
\begin{align}
\Phi^f_\pm = \pm \varphi^f_c +\Phi^f_s, \hspace{10pt} \Phi^g_\pm = \pm \varphi^g_c +\Phi^g_s. 
\end{align}
They satisfy the commutation relations, 
\begin{align}
\begin{split}
[\Phi^I_{\sigma,j}(x), \Phi^I_{\sigma,j'}(x')] &= i\pi I (-1)^j \delta_{jj'} \textrm{sgn} (x-x') +i\pi (1-\delta_{jj'}) \textrm{sgn} (j-j'), \\
[\Phi^I_{\sigma,j}(x), \Phi^I_{-\sigma,j'}(x')] &= i\pi \sigma, \\
[\Phi^I_{\sigma,j}(x), \Phi^{-I}_{\sigma,j'}(x')] &= i\pi I (-1)^j \delta_{jj'} +i\pi (1-\delta_{jj'}) \textrm{sgn}(j-j'), \\
[\Phi^I_{\sigma,j}(x), \Phi^{-I}_{-\sigma,j'}(x')] &= i\pi \sigma, 
\end{split}
\end{align}
with the identification $I=f/g=+/-$. 
Since all the commutators are $i\pi$ mod $2i\pi$, we can define the fermionic fields by 
\begin{align}
\psi^f_{\pm,j}(t,x) = \frac{1}{\sqrt{2\pi d_x}} e^{-i\Phi^f_{\pm,j}(t,x)}, \hspace{10pt}
\psi^g_{\pm,j}(t,x) = \frac{1}{\sqrt{2\pi d_x}} e^{-i\Phi^g_{\pm,j}(t,x)}, 
\end{align}
where $d_x$ is a short-distance cutoff. 
They properly satisfy anti-commutation relations. 
In terms of these fermionic fields, the actions $S^D_\textrm{wire}$ and $S_\textrm{tun}$ is rewritten as 
\begin{align}
S^D_\textrm{wire}[\psi^{f/g}_\pm, A^c_\mu, a^{f/g}_\mu] =&\ \int dt dx \sum_{\sigma=\pm} \biggl[ (\psi^f_\sigma)^\dagger i \bigl\{ \partial_t -ia^f_0 -i\sigma A^c_0 -vP (\partial_x -ia^f_1 -i\sigma A^c_1) \bigr\} \psi^f_\sigma \nonumber \\
&+(\psi^g_\sigma)^\dagger i \bigl\{ \partial_t -ia^g_0 -i\sigma A^c_0 +vP (\partial_x -ia^g_1 -i\sigma A^c_1) \bigr\} \psi^g_\sigma \biggr], \\
S_\textrm{tun}[\psi^{f/g}_\pm, \phi_j] =& \ -\int dt dx \sum_{j=0}^{N-1} \sum_{\sigma=\pm} \Bigl[ 2\pi d_x \Bigl\{ \left( g_1 -(-1)^j g_2 \phi_j \right) i (\psi^f_{\sigma,j})^\dagger \psi^f_{\sigma,j+1} \nonumber \\ 
&+\left( g_1 +(-1)^j g_2 \phi_j \right) i (\psi^g_{\sigma,j})^\dagger \psi^g_{\sigma,j+1} +\textrm{h.c.} \Bigr\} -c_1 \phi_j^2 +c_2 (\phi_j-\phi_{j+1})^2 +\cdots \Bigr].
\end{align}

We now take the continuum limit in the direction perpendicular to wires by introducing the continuous variable $y=2nd_y$ with $d_y$ being the lattice spacing between wires. 
Supposing that $g_2 \ll g_1$ and $c_1$ is close to zero, we can define the spinor fields by 
\begin{align}
\begin{split}
\frac{1}{\sqrt{2d_y}} \left( \begin{array}{cc} \psi^f_{\sigma,2n+1}(t,x) \\ \psi^f_{\sigma,2n}(t,x) \end{array} \right) &\sim \left( \begin{array}{cc} (\psi^f_\sigma)_1(t,x,y) \\ (\psi^f_\sigma)_2(t,x,y) \end{array} \right) \equiv \psi^f_\sigma(t,x,y), \\
\frac{1}{\sqrt{2d_y}} \left( \begin{array}{cc} \psi^g_{\sigma,2n+1}(t,x) \\ \psi^g_{\sigma,2n}(t,x) \end{array} \right) &\sim \left( \begin{array}{cc} (\psi^g_\sigma)_1(t,x,y) \\ (\psi^g_\sigma)_2(t,x,y) \end{array} \right) \equiv \psi^g_\sigma(t,x,y),
\end{split}
\end{align}
We similarly replace the gauge fields $a^{f/g}_{\nu,j}(t,x)$ and $A^{c/s}_{\nu,j}(t,x)$ ($\nu=0,1$) as well as the order-parameter field $\phi_j(t,x)$ by the slowly varying fields $a^{f/g}_\nu(t,x,y)$, $A^{c/s}_\nu(t,x,y)$, and $\phi(t,x,y)$, respectively. 
The total action is then given by 
\begin{align}
S_\textrm{dual}[\psi^{f/g}_\pm, A^{c/s}_\mu, a^{f/g}_\mu, \phi] \sim& \ \int dt dx dy \Biggl[ \sum_{\sigma=\pm} \bar\psi^{f}_\sigma  iw_\mu \gamma^\mu (\partial_\mu-ia^f_\mu -i\sigma A^c_\mu) \psi^f_\sigma +\sum_{\sigma=\pm} \bar\psi^g_\sigma iw_\mu \gamma^\mu (\partial_\mu -ia^g_\mu -i\sigma A^c_\mu) \psi^g_\sigma \nonumber \\
&+4\pi g_2 d_x \phi \sum_{\sigma=\pm} (\bar \psi^f_\sigma \psi^f_\sigma +\bar \psi^g_\sigma \psi^g_\sigma) +\frac{c_1}{d_y} \phi^2 -c_2 d_y (\partial_y \phi)^2 +\frac{1}{2\pi} \epsilon_{\mu \nu \rho} A^s_\mu \partial_\nu (a^f_\rho -a^g_\rho) \nonumber \\
&+\frac{1}{8\pi} \lambda_\mu (\epsilon_{\mu \nu \rho} \partial_\nu a^f_\rho)^2 +\frac{1}{8\pi} \lambda_\mu (\epsilon_{\mu \nu \rho} \partial_\nu a^g_\rho)^2 +\frac{1}{4\pi} \lambda_\mu (\epsilon_{\mu \nu \rho} \partial_\nu A^s_\rho)^2 +\cdots \Biggr],
\end{align}
where
\begin{align}
\begin{split}
(\gamma^0,\gamma^1,\gamma^2) &= (\sigma^y, -i\sigma^x, -i\sigma^z), \\
\bar\psi^{f}_\sigma &=(\psi^f_\sigma)^\dagger \gamma^0, \\
\bar\psi^{g}_\sigma &=(\psi^g_\sigma)^\dagger \gamma^0, \\
(w_0,w_1,w_2) &= (1, v, 4\pi g_1 d_x d_y), \\
(\lambda_0,\lambda_1,\lambda_2) &= (d_y/v,vd_y,0). 
\end{split}
\end{align}
Replacing $A^s_\mu \to -A^s_\mu$, we obtain the spatially anisotropic version of the action in the main text with the gauge $A^{c/s}_2=0$ and $a^{f/g}_2=0$. 

\section{$U(1)$ SPT, superfluid, anyon superfluid, trivial Mott insulator and their transition}
In general, we can modify our theory  to describe the effective theory of the $U(1)$ SPT, superfluid, anyon superfluid, trivial Mott insulator and their transition.
To obtain such theory, we don't couple the Dirac fermions to a bosonic field, instead we add the mass terms explicitly and get the Lagrangian,
\begin{align} \label{eq:QED_all}
  \mathcal L=&\sum_{\sigma=\pm}\bar \psi^f_\sigma [i\gamma^\mu(\partial_\mu-ia^f_\mu-i\sigma A^c_\mu) ]\psi^f_\sigma -\frac{1}{2\pi}\mcs{A^s}{a^f} \nonumber \\
+&\sum_{\sigma=\pm}\bar \psi^g_\sigma [i\gamma^\mu(\partial_\mu-ia^g_\mu-i\sigma A^c_\mu) ]\psi^g_\sigma +\frac{1}{2\pi}\mcs{A^s}{a^g} \nonumber \\
+&\sum_{\sigma=\pm}(m^f_\sigma \bar \psi^f_\sigma \psi^f_\sigma+ m^g_\sigma \bar \psi^g_\sigma \psi^g_\sigma) 
%-2\lambda \phi^2+ u \phi^4+\cdots,
%+&\sum_{t=\pm}(m^f_t\bar f_t f_t+m^g_t\bar g_t g_t)+\cdots.
\end{align}
By assigning different signs to the mass terms $m^f_\pm$ and $m^g_\pm$, we can get different phases; and the critical point can be achieved by tuning some mass to $0$. 

Let us start with  describing different phases by choosing different masses for the Dirac fermions,
\begin{enumerate}
\item $U(1)$ SPT phase: positive $U(1)$ SPT ($\sigma_c^{xy}=2$), $m^f_\pm, m^g_\pm>0$; negative $U(1)$ SPT ($\sigma^{xy}_c=-2$) $m^f_\pm, m^g_\pm<0$.
As we show before, it can be understood by integrating out the gapped Dirac fermions that yields a Hall response term,
\begin{align}
\mathcal L=&-\frac{2}{4\pi}\textrm{sgn} (m) \vp[ \scs{A^c}-\scs{A^s}].
\end{align}
Note that the time-reversal symmetry is explicitly broken here, which is different from the case we discussed in the paper. 
\item Trivial Mott insulating phase: $m_\pm^f>0,m_\pm^g<0$ (or  $m_\pm^f<0,m_\pm^g>0$). Integrating out the Dirac fermions will give
 \begin{align}
\mathcal L=&-\frac{1}{4\pi} \vp[ \scs{A^c}+\scs{a^f}]-\frac{1}{2\pi}\mcs{A^s}{a^f} \nonumber \\
		  & +\frac{1}{4\pi} \vp[\scs{A^c}+\scs{a^g}]+\frac{1}{2\pi}\mcs{A^s}{a^g}.	 % &+\frac{1}{4e^2} \mw{a^f}+\frac{1}{4e^2} \mw{a^g}.
\end{align}
Similar as the $U(1)$ SPT,  Chern-Simons terms for the internal gauge fields $a^f$ and $a^g$ are generated, hence the photons of those gauge fields are gapped. Therefore, the resultant state is a Mott insulating phase, but compared with the $U(1)$ SPT phase, the Hall response here is $0$. This can be seen by integrating out $a^f$ and $a^g$,
 \begin{align}
\mathcal L=&-\frac{1}{4\pi} \vp[ \scs{A^c}-\scs{A^s}] \nonumber \\
		  & +\frac{1}{4\pi} \vp[\scs{A^c}-\scs{A^s}].	 % &+\frac{1}{4e^2} \mw{a^f}+\frac{1}{4e^2} \mw{a^g}.
\end{align}
Now it is clear that the Chern-Simons term generated by Dirac fermions  $\psi^f_\pm$ and $\psi^g_\pm$  cancels each other. 
\item Superfluid phase: $m^f_+>0, m^f_-<0$, $m^g_+>0, m^g_-<0$ (or other similar combinations). For such cases, integrating out Dirac fermions will give
\begin{align}
\mathcal L=-\frac{1}{2\pi}\vp\smcs{A^c}{(a^f+a^g)}-\frac{1}{2\pi}\vp\smcs{A^s}{(a^f-a^g)}.
\end{align}
As we argued before, it describes a superfluid phase.
\item Anyon superfluid phase: $m^f_+>0, m^f_-<0$, $m^g_\pm>0$ (or other similar combinations). Integrating out Dirac fermions will give,
\begin{equation}
\mathcal L=-\frac{1}{2\pi}\mcs{A^c}{a^f}-\frac{1}{2\pi}\mcs{A^s}{a^f}-\frac{1}{4\pi} \vp[ \scs{A^c}-\scs{A^s}]
\end{equation}
Again it is a superfluid phase, as the photon of the internal gauge field $a^f$ can be identified as the goldstone mode. However, compared with the previous case, there is indeed a quantized Hall response for the superfluid phase.
\end{enumerate}

Now we can easily obtain the theory of the phase transition  between those different phases, some of which are discussed in Ref. \cite{Lu2014, Grover2013}.
For example, the transition from the $U(1)$ SPT ($m^f_\pm,m^g_\pm>0$) to the trivial Mott insulating phase ($m^f_\pm>0,m^g_\pm<0$), can be is described by $m^g_\pm=0$ and $m^f_\pm>0$:
\begin{align}
\mathcal L&=\sum_{\sigma=\pm}\bar \psi^f_\sigma [i\gamma^\mu(\partial_\mu-ia^f_\mu-i\sigma A^c_\mu) ]\psi^f_\sigma -\frac{1}{2\pi}\mcs{A^s}{a^f}   -\frac{1}{4\pi} \vp[ \scs{A^c}-\scs{A^s}].
\end{align}
On the other hand, the transition from the $U(1)$ SPT ($m^f_\pm,m^g_\pm>0$) to the anyon superfluid ($m^f_+>0, m^f_-<0$, $m^g_\pm>0$) can be achieved by tuning one Dirac mass (say $m^f_-$), and its effective theory is
\begin{equation}
\mathcal L=\bar \psi^f_- [i\gamma^\mu(\partial_\mu-ia^f_\mu+iA^c_\mu) ] \psi^f_- -\frac{1}{2\pi}\mcs{A^s}{a^f}-\frac{1}{8\pi} \cs{(a^f+A^c)}   -\frac{1}{4\pi} \vp[ \scs{A^c}-\scs{A^s}].
\end{equation}

Finally the transition between the opposite $U(1)$ SPT is described by tuning four Dirac masses simultaneously, and its effective theory is
\begin{align} \label{eq:op_SPT}
  \mathcal L=&\sum_{\sigma=\pm}\bar \psi^f_\sigma [i\gamma^\mu(\partial_\mu-ia^f_\mu-i\sigma A^c_\mu) ]\psi^f_\sigma -\frac{1}{2\pi}\mcs{A^s}{a^f} \nonumber \\
+&\sum_{\sigma=\pm}\bar \psi^g_\sigma [i\gamma^\mu(\partial_\mu-ia^g_\mu-i\sigma A^c_\mu) ]\psi^g_\sigma +\frac{1}{2\pi}\mcs{A^s}{a^g}+\cdots
\end{align}
This is similar as the transition between the superfluid and the $U(1)$ SPT with the spontaneous symmetry breaking that we described in the paper.

One may realize that most of those transitions will generically require the tuning of more than one masses, hence seem to be fine tuned.
One  way to avoid such fine tuning is using certain symmetry to enforce the Dirac masses changing simultaneously. 
However we want to emphasize that, the phase transition described in the paper, namely the transition between the superfluid phase and the $U(1)$ SPT phase with spontaneous time-reversal symmetry breaking, is not fine tuned. 
The reason is because in this case, the Dirac masses are dynamically generated from the spontaneous symmetry breaking, hence are guaranteed to change simultaneously. 

\section{The critical theory for the general coupling constant}

The Lagrangian for the kagome spin liquid is
\begin{eqnarray}
\mathcal{L}&=&\sum_{\sigma=\pm}\bar \psi^f_\sigma\gamma^\mu\left(\partial_\mu-ia^f_\mu-i\frac{\sigma A_{\rm ext}}{2}\right)\psi^f_\sigma+\sum_{\sigma=\pm}\bar \psi^g_\sigma\gamma^\mu\left(\partial_\mu-ia^g_\mu-i\frac{\sigma A_{\rm ext}}{2}\right)\psi^g_\sigma\nonumber\\
&&-\frac{\epsilon^{\mu\nu\lambda}}{2\pi}\mathcal{A}_\mu\partial_\nu(a^f_\lambda-a^g_\lambda)+\frac{1}{e^2}\left[(f^{(f)}_{\mu\nu})^2+(f^{(g)}_{\mu\nu})^2\right]+\frac{1}{e_0^2}(\mathcal{F}_{\mu\nu})^2.
\end{eqnarray}
Here and in the following, $(f_{\mu\nu})^2$ represents the Maxwell term for the dynamical gauge field.
We can drop $A_{\rm ext}$ (probe field) for the rest of this argument to get
\begin{eqnarray}
\mathcal{L}&=&\sum_{\sigma=\pm}\bar \psi^f_\sigma\gamma^\mu\left(\partial_\mu-ia^f_\mu \right)\psi^f_\sigma+\sum_{\sigma=\pm}\bar \psi^g_\sigma\gamma^\mu\left(\partial_\mu-ia^g_\mu\right)\psi^g_\sigma
%\sum_{t=\pm}\bar f_t\gamma^\mu\left(\partial_\mu-ia^f_\mu\right)f_t+\sum_{t=\pm}\bar g_t\gamma^\mu\left(\partial_\mu-ia^g_\mu\right)g_t
-\frac{\epsilon^{\mu\nu\lambda}}{2\pi}\mathcal{A}_\mu\partial_\nu(a^f_\lambda-a^g_\lambda)\nonumber\\
&&+\frac{1}{e^2}\left[(f^{(f)}_{\mu\nu})^2+(f^{(g)}_{\mu\nu})^2\right]+\frac{1}{e_0^2}(\mathcal{F}_{\mu\nu})^2.
\end{eqnarray}
Now define 
\begin{eqnarray}
a^\pm=a^f\pm a^g
\end{eqnarray}
to get
\begin{eqnarray}
\mathcal{L}&=&\sum_{\sigma=\pm}\bar \psi^f_\sigma\gamma^\mu\left(\partial_\mu-i\frac{a^+_\mu+a^-_\mu}{2}_\mu\right)\psi^f_\sigma+\sum_{\sigma=\pm}\bar \psi^g_\sigma\gamma^\mu\left(\partial_\mu-i\frac{a^+_\mu-a^-_\mu}{2}\right)\psi^g_\sigma-\frac{\epsilon^{\mu\nu\lambda}}{2\pi}\mathcal{A}_\mu\partial_\nu a^-_\lambda\nonumber\\
&&+\frac{1}{4e^2}\left[(f^{(+)}_{\mu\nu})^2+(f^{(-)}_{\mu\nu})^2\right]+\frac{1}{e_0^2}(\mathcal{F}_{\mu\nu})^2
\end{eqnarray}
which we can write as:
\begin{eqnarray}
\mathcal{L}&=&\sum_{\sigma=\pm}\bar \psi^f_\sigma\gamma^\mu\left(\partial_\mu-i\frac{a^+_\mu+a^-_\mu}{2}_\mu\right)\psi^f_\sigma+\sum_{\sigma=\pm}\bar \psi^g_\sigma\gamma^\mu\left(\partial_\mu-i\frac{a^+_\mu-a^-_\mu}{2}\right)\psi^g_\sigma-\frac{\epsilon^{\mu\nu\lambda}}{4\pi}\left[\mathcal{A}_\mu\partial_\nu a^-_\lambda-a^-_\mu\partial_\nu\mathcal{A}_\lambda\right]\nonumber\\
&&+\frac{1}{4e^2}\left[(f^{(+)}_{\mu\nu})^2+(f^{(-)}_{\mu\nu})^2\right]+\frac{1}{e_0^2}(\mathcal{F}_{\mu\nu})^2
\end{eqnarray}
Now let us define
\begin{eqnarray}
\alpha=\mathcal{A}+ a^-~~~~~~~~\beta=\mathcal{A}-a^-
\end{eqnarray}
to get
\begin{eqnarray}
\mathcal{L}&=&\sum_{\sigma=\pm}\bar \psi^f_\sigma\gamma^\mu\left(\partial_\mu-i\frac{a^+_\mu}{2}-i\frac{\alpha_\mu-\beta_\mu}{4}\right)\psi^f_\sigma+\sum_{\sigma=\pm}\bar \psi^g_\sigma\gamma^\mu\left(\partial_\mu-i\frac{a^+_\mu}{2}+i\frac{\alpha_\mu-\beta_\mu}{4}\right)\psi^g_\sigma-\frac{2\epsilon^{\mu\nu\lambda}}{4\pi}\left[\alpha_\mu\partial_\nu \alpha_\lambda-\beta_\mu\partial_\nu\beta_\lambda\right]\nonumber\\
&&+\frac{1}{4e^2}(f^{(+)}_{\mu\nu})^2+\left[\frac{1}{16e^2}+\frac{1}{4e_0^2}\right]\left[(f^{(\alpha)}_{\mu\nu})^2+(f^{(\beta)}_{\mu\nu})^2\right]+\left[\frac{1}{8e^2}-\frac{1}{2e_0^2}\right]f^{(\alpha)}_{\mu\nu} f^{(\beta)}_{\mu\nu}
\end{eqnarray}
Extremising for the equations of motions of $\alpha$ and $\beta$ for the pure gauge sector, we get:
\begin{eqnarray}
\left[\begin{array}{cc}
\frac{1}{16e^2}+\frac{1}{4e_0^2} & \frac{1}{8e^2}-\frac{1}{2e_0^2}\\
\frac{1}{8e^2}-\frac{1}{2e_0^2} & \frac{1}{16e^2}+\frac{1}{4e_0^2}\\
\end{array}\right]\left[\begin{array}{c}
\partial_\mu f^{(\alpha)}_{\mu\nu}\\
\partial_\mu f^{(\beta)}_{\mu\nu}\\
\end{array}\right]=\frac{\epsilon^{\nu\lambda\rho}}{\pi}\left[\begin{array}{cc}
1 & 0\\
0 & -1\\
\end{array}\right]\left[\begin{array}{c}
f^{(\alpha)}_{\lambda\rho}\\
f^{(\beta)}_{\lambda\rho}\\
\end{array}\right]
\end{eqnarray}
For general values of $e$ and $e_0$, both the photons corresponding to $\alpha$ and $\beta$ will aquire a mass and hence can be integrated out. This means we can drop the corresponding couplings (up to irrelevant short range four fermi interactions) to get the critical theory as:
\begin{eqnarray}
\mathcal{L}&=&\sum_{\sigma=\pm}\bar \psi^f_\sigma\gamma^\mu\left(\partial_\mu-i\frac{a^+_\mu}{2}\right)\psi^f_\sigma+\sum_{\sigma=\pm}\bar \psi^g_\sigma\gamma^\mu\left(\partial_\mu-i\frac{a^+_\mu}{2}\right)\psi^g_\sigma
\end{eqnarray}
Which is the same theory obtained in the large $e_0$ limit. This, then is the critical theory.

\end{document}